\documentclass[epjCONF, onecolumn]{svjour}
\usepackage{graphics}
\usepackage{bm}
\usepackage[varg]{txfonts} 
\usepackage[latin1]{inputenc}

\newcommand{\be}{\begin{eqnarray}}
\newcommand{\ee}{\end{eqnarray}}

\newcommand{\del}{\partial}
\newcommand{\tr}{\, {\rm tr}\, }

\session-title{Hadron Nuclear Physics (HNP) 2011}
\begin{document}
\title{Hadron resonances with coexistence of different natures}
\author{Atsushi Hosaka\thanks{\email{hosaka@rcnp.osaka-u.ac.jp}}$^{1}$, 
Tetsuo Hyodo$^{2}$, 
Daisuke Jido$^{3}$, 
Hideko Nagahiro$^{4}$, 
Kanabu Nawa$^{1}$, 
Shunsuke Ohkoda$^{1}$, 
Sho Ozaki$^{5}$,
Yasuhiro Yamaguchi$^{1}$ and 
Shigehiro Yasui$^{6}$}
\institute{
$^1$Research Center for Nuclear Physics (RCNP), Osaka University, Ibaraki, 567-0047, Japan\\
$^2$Department of Physics, Tokyo Institute of Technology, Meguro 152-8551, Japan\\
$^3$Yukawa Institute for Theoretical Physics, Kyoto University, Kyoto 606-8502, Japan\\
$^4$Department of Physics, Nara Women's University, Nara 630-8506, Japan\\
$^5$Department of Physics, The University of Tokyo, Tokyo 113-0033, Japan\\
$^6$KEK Theory Center, Institute of Particle and Nuclear Studies, High Energy Accelerator Research Organization,
1-1, Oho, Ibaraki, 305-0801, Japan
}
\abstract{We discuss coexistence/mixing of different natures of hadronic composite (molecule) and elementary (quark-intrinsic) ones in hadron resonances.  
The discussions here are 
based on our previous publications on the origin of hadron resonances~\cite{Hyodo:2008xr}, 
exotic $\bar D$ meson-nucleons as hadronic composites containing one anti-heavy quark~\cite{Yamaguchi:2011xb}, and the study of $a_1$ as a typical example to show explicitly the mixing of the two different natures~\cite{Nagahiro:2011jn}.  
In all cases, interactions are derived from the chiral dynamics of the light flavor sector.  
These interactions generate in various cases hadronic composite/molecule states, 
serving varieties of structure beyond the conventional quark model.  }
\maketitle

\section{Introduction}

One of recent activities in hadron physics has been motivated  
by the observations of exotic hadrons.  
The candidate of the truely exotic five quark state $\Theta^+$ is now 
under further investigations~\cite{Nakano:2003qx,Nakano:2008ee}, 
while the existence and its nature of 
$X(3872)$ is being  established~\cite{Bhardwaj:2011dj}.   
Very recently yet strong candidates
of the true tetraquark states in the bottom sector have been 
observed near the $B\bar B^*$ ($B^* \bar B$) 
and $B^* \bar B^*$ thresholds~\cite{Collaboration:2011gja}.  

In general, multiquark components are expected to exist 
also in the conventional hadron resonances.  
Because a typical excitation energy of hadron resonances 
is of order  half GeV,  additional pair of quark and anti-quark 
may be created, forming a multiquark component in the resonance state.  
A multiquark component may rearrange itself such that its energy takes 
the minimum.  
Among such correlations  diquark, triquark and hadronic ones are 
widely assumed~\cite{Jaffe:2003sg,Karliner:2003dt}.  
The former two form colored clusters existing only inside hadrons, 
while the latter may develop a molecular like structure; 
two (or more in general) hadrons are loosely bound while 
each hadron keeps its identity.  
In other words, such states have a larger spatial size as compared to 
their constituent hadrons.  

Here we consider several hadronic molecules, 
or in what follows we shall call them hadronic composite, because we
know  their interactions better than the interaction between colored clusters.  
We expect that  such states should be seen  
near the threshold region of the constituent hadrons, 
though their appearance is not trivial, depending on the nature 
of the relevant hadron interactions.  
If the interaction is too strong, they may form a strongly bound 
system which cannot keep the identity of a constituent hadron. 

The hadronic composite is another candidate of exotic hadrons 
in wider sense, although we 
have not yet confirmed any example of such hadronic composites 
except for atomic nuclei.  
At  present, almost all hadrons listed in the particle data
are explained regarding their quantum numbers by the constituent 
quark model where mesons and baryons are described by $q \bar q$ and 
$qqq$, respectively.   
It is, however, widely expected that some of hadrons have 
different structure from those of the constituent quark model, 
such as $\Lambda(1405)$~\cite{Kaiser:1995eg,Oset:1997it}
and scalar mesons  
$\sigma(600), f_0(980), a_0(980)$~\cite{Oller:1997ti,Pelaez:2003dy}.  
They are rather generated by coupled channels of  
$\bar K N$-$\pi \Sigma$~\cite{Jido:2003cb,Hyodo:2007jq} 
and $\pi \pi$-$\bar KK$, respectively.  

Once we have such hadronic descriptions with reliable 
interactions, we solve the scattering problem, 
where resonance states appear as  poles of the scattering amplitude. 
This is how hadronic composites are described and are 
dynamically generated.  

In general, however, we need  explicit introduction of intrinsic 
degrees of freedom which is not in the original theory of hadronic 
composites to explain experimental data.  
This is natural because many resonances are 
described by the quark model.  
In the examples we discuss here, the necessity is associated 
with the structure of the constituent hadrons having a smaller
size than the hadronic composites.  
Therefore, we assume two hierarchies discriminated by spatial size; 
hadrons described by the quark model with a smaller size
and hadronic composites with a larger size.  
Following Weinberg~\cite{Weinberg:1962hj}, we call the quark model states elementary, 
in the sense that they are not easily described as composite states of 
hadrons.  

In this report, we discuss three topics.  
One is the introduction of the elementary component in a 
hadronic model based on the scattering of a Nambu-Goldstone boson 
in the chiral unitary approach~\cite{Hyodo:2008xr}.  
The interaction for the Nambu-Goldstone bosons 
is given by the Weinberg-Tomozawa theorem.  
This picture has been extensively studied for kaon dynamics 
which seems to have an ideal ground to form  
hadronic composites.  
Second we discuss hadronic composites in the heavy quark sector 
where we predict several bound and resonant states 
of $\bar D$ and $B$ mesons~\cite{Yamaguchi:2011xb,Yasui:2009bz}. 
The interaction is mediated by the pion, which becomes 
crucial when the coupled channel effect becomes important 
though the mixing of S and D waves due to the tensor 
component of the pion exchange force.  
The mechanism is similar to the one for the deuteron 
where the one pion exchange force plays a crucial role 
for the binding of the deuteron~\cite{ericsonweise}.  
For the $\bar D$ and $B$ meson cases, the degeneracy between 
$\bar D$-$\bar D^*$ and $B$-$B^*$ mesons are important, which is 
indeed the case in the limit of heavy quark mass.  
In the third topic we discuss the coexistence of the
two different structures, the hadronic composite  
and elementary components~\cite{Nagahiro:2011jn}.  
An analysis is made for the axial vector meson $a_1$ where 
well established chiral Lagrangians with vector mesons are available.  

\section{Hadronic composites in the chiral unitary model}

The chiral unitary model has been extensively used for the study 
of hadron resonances as dynamically generated 
states~\cite{Oset:1997it,Jido:2003cb,Hyodo:2007jq,Oller:1997ti,Pelaez:2003dy}.  
It is based on the chiral perturbation theory and the  
basic amplitudes at low energies are 
provided by the Weiberg-Tomozawa (WT) interaction
for the S-wave scattering of a Nambu-Goldstone boson from a baryon 
(in general, matter).  
This low energy amplitude is then 
unitarized to extend into the resonance region.  
Typical successful cases are for $\Lambda(1405)$ and 
scalar mesons
as resonances as 
$\bar K N$-$\pi\Sigma$~\cite{Jido:2003cb,Hyodo:2007jq} 
and $\pi \pi$-$\bar{K} K$
scattering.  
The WT amplitude is given by the Lagrangian at the tree level
\be
L_{WT} = 
\frac{1}{f_\pi^2} V^\mu_c f_{abc} (\del_\mu \phi^a) \phi^b
\label{MBvertex}
\ee
where $f_\pi$ is the pion decay constant, $f_{abc}$ are the structure 
constant of flavor SU(3).  
The octet NG boson field is denoted by $\phi^a$, and $V^\mu_c$ are for 
the octet components of the vector current of the target particle.  

The amplitude is then unitarized by using (\ref{MBvertex}) 
as a potential for the Schr\"odinger equation.
We can solve it and derive  
the scattering matrix $T$; 
\be
T(E) = V_{WT} + V_{WT}G(E) T\, 
\label{LSeq}
\ee
where the Weinberg-Tomozawa interaction $V_{WT}$
is obtained from the Lagrangian 
(\ref{MBvertex}).
By using the pointlike (separable) nature of the interaction $V_{WT}$ as well 
as the on-shell factorization~\cite{Oset:1997it}, 
the two-body propagator $G(E)$ is separated from the integral equation, 
and is given in the form of 
a one-loop integral, which can be performed analytically.  
However, precisely due to the pointlike nature, the loop 
integral diverges and a suitable renormalization is needed to obtain 
a finite result.  
Two methods are often used, one is the dimensional regularization
with removing the divergent  terms and introducing a subtraction
(renormalization) constant $a$.  
Another is to express it as an integral 
over three dimensional momentum and 
perform the integral up to a cutoff momentum $\Lambda$,  
which is equivalent to the sum over intermediate states
in the second order perturbation theory:
\be
G(E) \sim \sum_n \frac{1}{E - E_n}
\label{G_sum_n}
\ee
Mathematically, this sum runs up to infinitely large momentum, while 
a finite size structure of the constituent hadrons 
may render a cutoff, giving a finite result.  
In this case due to its structure of (\ref{G_sum_n}), 
the G-function takes a negative value 
for energies below the threshold ($E < E_{n}$ for all $n$).

In Ref.~\cite{Hyodo:2008xr}, this property was used to introduce the natural 
condition for the G-function calculated in the dimensional regularization, 
which determines the natural subtraction constant $a = a_{\rm natural}$.  
If resonance properties are reproduced by using this natural value, 
we can say that the resonance is dynamically generated and is described 
as a hadronic composite.  
It was shown  in Ref.~\cite{Hyodo:2008xr} that
$\Lambda(1405)$ is a typical resonance of such.  
The natural value $a_{\rm natural}$ for this case corresponds to 
the cutoff $\Lambda \sim 600$ MeV, consistent with 
$\bar K N$ (loosely) bound state with  intrinsic size of constituent 
hadrons of order of 0.5 fm.  

In many cases, however, resonance properties are not reproduced 
by using the natural subtraction constant, 
where we need to fix it at 
$a \neq a_{\rm natural}$.  
It was shown that this was indeed the case for $N(1535)$~\cite{Hyodo:2008xr}.  
What is interesting, however, that when 
$a \neq a_{\rm natural}$, 
the difference 
$\Delta a = a - a_{\rm natural}$
introduces a pole-like interaction additionally: 
\be
T(E) = \frac{1}{V_{WT}^{-1} - G(E; a)} 
= \frac{1}{V_{\rm eff}^{-1} - G(E; a_{\rm natural})} 
\ee
where
\be
V_{\rm eff} = V_{WT} + \frac{C}{2f_\pi^2}
\frac{(E-M)^2}{E- M_{\rm eff}^2},  \; \; \; 
M_{\rm eff} = M - \frac{16 \pi^2 f_\pi^2}{CM\Delta a}
\ee
In these equations, $M$ denotes the mass of the target particle, 
and $C$ is a constant for the Weinberg-Tomozawa
interaction.  
The additional piece in the interaction may be then interpreted 
as the propagation of an elementary particle which is not in the original 
model space of the two-body scattering, 
just like the CDD pole~\cite{Castillejo:1956ed}.  
In this way, we can see that resonances, in general, are 
mixture of the hadronic composite and elementary components.  
More detailed analysis for the  compositeness or elementariness 
of a particle is performed in Ref.~\cite{Hyodo:2011qc}.

\section{$\bar D$ molecule as a pentaquark baryon}

As a good example of hadronic composite, we discuss in this section heavy 
quark systems of $\bar D N$ and  $B N$.  
The minimal quark content of these states is 
$\bar Q qqqq$ ($Q = b, c; q = u, d$), and therefore, they are analogues 
of the pentaquark $\Theta^+ \sim \bar s uudd$ with one heavy 
anti-quark.  
An interesting feature in the heavy quark sector is the heavy quark 
symmetry which leads to the degeneracy of spin 0 and 1 particle; 
$\bar D$ and $\bar D^*$, and $B$ and $B^*$.  

\begin{figure}[h]
\resizebox{0.4\columnwidth}{!}{%
  \includegraphics{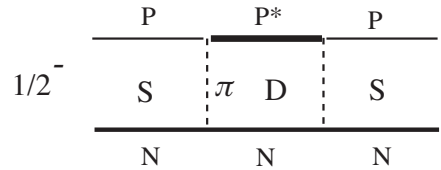} }
\centering
\begin{minipage}{13cm}
\caption{\small Pion exchange between the $P$ ($P^*$) meson and the nucleon ($N$).}
\label{opep}
\end{minipage}
\end{figure}

From now on let us introduce the notation  $P$  for the pseudoscalar meson, 
and $P^*$ for the vector meson.  
Because of the degeneracy, the channel coupling of $P^*N$ and $PN$ 
becomes important.  
This contrasts to the light flavor system, where the mass difference
of $K^*$ and $K$, and that of $\rho$ and $\pi$ are much larger than 
that for the charm and bottom sectors, as summarized in Table~\ref{masses}.
In this case the pion coupling at the vertex $P^*P\pi$ and $NN\pi$ allows 
one pion exchange potential as shown in Fig.~\ref{opep}.  
A characteristic feature of the one-pion exchange is then the 
tensor force as in the nuclear system; for instance, it is well known 
that the tensor force causes the channel coupling of S and D waves 
leading to a strong attraction for the deuteron.  
The same mechanism applies to the heavy meson system
through the transition $P^*N \to PN$.  
We note that the $P^*P\pi$ coupling has the structure of 
$\vec s \cdot \vec q$ where $\vec s$ is a (half of) spin operator and $\vec q$
the momentum carried by the pion. 
For the case of $NN\pi$, $\vec s$ is the Pauli matrix, while for $P^*P\pi$
it is the spin transition operator between spin 1 and 0 states.  

\begin{table}[h]
\begin{center}
\caption{Mass differencees of pseudoscalar and vector mesons.}
\begin{tabular}{cccc}
\hline 
$m_\rho - m_\pi$  & $m_{K^*} - m_K$ 
   & $m_{\bar D^*} - m_{\bar D} $ & $m_{B*} - m_B$ \\
\hline
630 MeV & 400 MeV & 140 MeV & 45 MeV\\
\hline
\end{tabular}
\label{masses}
\end{center}
\end{table}

 
The above idea has been proposed and tested in \cite{Yasui:2009bz}, 
and further elaborated in 
\cite{Yamaguchi:2011xb}.  
Then it turns out that precisely due to the  mixing mechanism, 
the system develops 
a low lying bound state, in this case in $J^P = 1/2^-$.
Furthermore, 
a resonance state was found for $J^P = 3/2^-$.  
The necessary coupled channels are shown in Table~\ref{coupledchannels}.  
In Ref.~\cite{Yamaguchi:2011xb}, a short range interaction mediated by 
$\rho$ and $\omega$ exchanges was also included, 
where it was shown 
that these meson exchanges play only minor role
as discussed below (Table~\ref{tab:bound}).  

\begin{table}[h]
\centering
\caption{\label{coupledchannels} \small Various coupled channels for a 
given quantum number $J^P$ for negative parity $P = -1$.  }
\vspace*{0.5cm}
{\small 
\begin{tabular}{ c  | c c c c}
\hline
$J^P$ &  \multicolumn{4}{c }{channels} \\
\hline
$1/2^-$ &  $PN(^2S_{1/2})$ & $P^*N(^2S_{1/2})$ & 
     $P^*N(^4D_{1/2})$ & \\
$3/2^-$ &  $PN(^2D_{3/2})$ & $P^*N(^2D_{3/2})$ & 
     $P^*N(^4D_{3/2})$ & $P^*N(^4S_{3/2})$ \\
\hline
\end{tabular}
}
\end{table}

The bound state properties are summarized in Table~\ref{tab:bound}
for the two cases when the one pion ($\pi$) and 
$\pi, \rho, \omega$ exchange potentials are employed.  
For the $\bar DN$ system, the binding energy and spatial size 
are similar to those of the deuteron.  
The system is then indeed loosely bound and consistent with 
a picture of hadronic composite. 
For the heavier system of the bottom quark, the binding energy becomes larger
due to stronger SD coupled channel effect and to less kinetic energy of heavier 
mass particle.  
Yet the size is larger than 1 fm, and we may regard them as hadronic 
composites.  

\begin{table}[h]
 \begin{center}
  \caption{The binding energies and the root mean
  square radii of the $J^P=1/2^-$ bound states.}
  \label{tab:bound}
  \begin{tabular}{l | cccc}
 \hline
 & $\bar{D}N(\pi)$ &  $\bar{D}N(\pi \rho \omega)$ & 
 $BN(\pi)$ &  $BN(\pi \rho \omega)$ \\ 
 \hline
Binding energy [MeV] & 1.60 & 2.14 & 19.50 & 23.04 \\
Root mean square radius [fm] & 3.5 & 3.2 & 1.3 & 1.2 \\
\hline
 \end{tabular}
 \end{center}
\end{table}

\begin{figure}[h]
\begin{tabular}{cc}
\resizebox{0.45\columnwidth}{!}{%
  \includegraphics{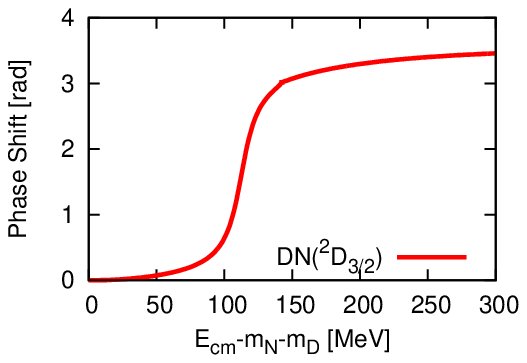} }
&
\resizebox{0.45\columnwidth}{!}{%
  \includegraphics{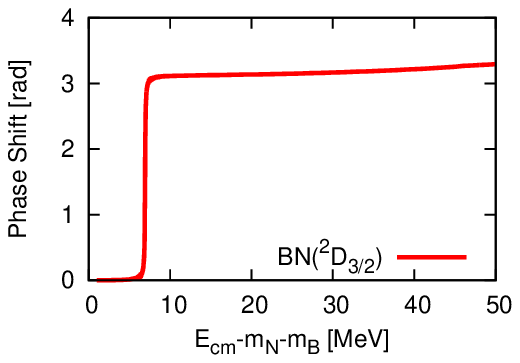} }
\end{tabular}
\centering
\begin{minipage}{13cm}
\caption{\small Phase shifts of the $\bar{D}N$ and $BN$ scattering  for
$J^P=3/2^-$ with $I=0$.}
\label{phaseshift}
\end{minipage}
\end{figure}

In Ref.~\cite{Yamaguchi:2011xb}, resonance states are also found for $J^P = 3/2^-$
for both charm and bottom sectors.  
This is the so called Feschbach resonance, where a would-be bound state
becomes a resonance when the coupling to the open 
$PN$ channel is turned on.  
In Fig.~\ref{phaseshift}, phase shifts of $PN$ scatterings are shown 
as functions of the scattering energy from the threshold.  
Phase shifts indeed cross at $\delta = \pi/2$ sharply indicating the presence 
of a narrow resonance.  
We have extracted resonance energies and the corresponding width as 
summarized in Table~\ref{tab:resonance}.  
The narrow width can be understood by the D-wave nature of the decay 
channel of $PN$ and small phase space volume.  

\begin{table}[h]
\caption{The resonance energy and decay width in the $J^P=3/2^-$ channel with $\pi,\rho,\omega$ exchange potential.}
\label{tab:resonance}
 \begin{center}
 \begin{tabular}{lccc}
 \hline 
   &$E_{re}-m_N-m_P$ [MeV]&$E_{re}$ [MeV] &Decay width [MeV] \\
\hline
$\bar{D}N$&113.19&2919.09&17.72 \\
$BN$&6.93&6224.83&$9.46\times 10^{-2}$ \\
\hline
  \end{tabular}
 \end{center}
\end{table}

\section{Coexistence of different natures in axial vector meson $a_1$}

In this section, we now discuss mixing of wave functions of different natures. 
The motivation was already discussed in the previous section 2, where the 
subtraction constant introduced an elementary component in the hadronic 
description of resonances.  
Here, we explicitly consider such a problem by using 
a suitable model for the hadronic description as well as for 
the elementary one.  

As an example, we study the axial vector meson $a_1(1260)$~\cite{Nagahiro:2011jn}.  
Conventionally $a_1$ has been considered as a chiral partner of 
the vector $\rho$ meson, which can be explicitly realized 
by a $\bar q q$ meson~\cite{Wakamatsu:1988ht,Hosaka:1990sj}.  
Recently, it has been also dynamically generated as a 
resonance of $\pi$-$\rho$ scattering and various properties 
have been investigated~\cite{Roca:2005nm,Nagahiro:2008cv}. 
In fact, the chiral Lagrangian of $\rho$ and $a_1$ contains 
the both aspects of $a_1$.  
For instance, one can employ 
the hidden local symmetry approach for the vector mesons~\cite{Bando:1984ej}, 
\be
L^{HLS}_{\pi \rho a_1} & = &
+ 2 f^2_{\pi } \tr  A^2_\mu  
-{1 \over {2}} \tr (  V^2_{\mu \nu } +  A^2_{\mu \nu })
\nonumber     \\
& + & 2 f^2_\pi g^2  \tr \left( V_\mu + \frac{1}{g} v_\mu \right)^2
+  2 f^2_\pi  g^2 \tr \left(  A_\mu + \frac{1}{g} a_\mu\right)^2
\label{L_HLS}
\ee
where the vector and axial vector currents of the pion field are defined by
\be
v_{\mu} =
\frac{i}{2} 
(\del_{\mu} \xi^\dagger \ \xi + \del_{\mu} \xi \ \xi^\dagger )\, , \; \; \; 
a_{\mu} =
\frac{i}{2} 
(\del_{\mu} \xi^\dagger \ \xi - \del_{\mu} \xi \ \xi^\dagger ) 
\label{pion_current}
\ee
\be
\xi = \exp(i\vec \tau \cdot \vec \pi/(2f_\pi)) \equiv \exp(i\pi/(2f_\pi) )
\ee
In Eq.~(\ref{L_HLS}) symbols $V$ and $A$ are for the $\rho$ 
and $a_1$ mesons, respectively.  
It is shown that  the Lagrangian (\ref{L_HLS}) can be derived from the 
extended NJL model, where the mesons are constructed by $\bar qq$, 
and are considered as elementary particles~\cite{Wakamatsu:1988ht,Hosaka:1990sj}.  
Recently,  a similar Lagrangian was derived from a holographic model 
of the string theory for QCD~\cite{Sakai:2004cn}. 

The Lagrangian contains sufficient ingredients for the present discussion of 
$a_1$; 
(1) elementary $a_1$ manifestly, (2) the coupling of the elementary 
$a_1$ to $\pi$ and $\rho$, and (3) the $\pi\rho$-$\pi\rho$ interaction for 
the dynamical generation of $a_1$ as a composite state.  
Hence,  now the problem is to solve a coupled channel system 
for the elementary $a_1$ and the $\pi\rho$ channels.  
In the literature, full flavor SU(3) coupled channels are 
included~\cite{Roca:2005nm,Nagahiro:2008cv}.
However, essential properties of $a_1$ are reproduced by the single channel of 
$\pi\rho$, and so we consider here only this channel.  

To proceed, it is convenient to introduce the Hamiltonian in the 
coupled channel form; 
\be
H = \left(
\begin{array}{cc}
H_{\pi\rho} + V_{WT} & g\\
g & H_e
\end{array}
\right)
\label{H22}
\ee
where $H_{\pi\rho}$ is the free $\pi\rho$ Hamiltonian, 
$H_{e}$ for the elementary $a_1$ which is simply its mass in the center of mass frame, and 
$g$ is the coupling of the elementary $a_1$ to $\pi\rho$.  
By introducing a two-component wave function 
$
\psi
= (\psi_{\pi\rho}, \psi_e)
$, 
we have coupled channel equations
\be
& & (H_{\pi\rho} + V_{WT}) \psi_{\pi\rho} + g \psi_e = E \psi_{\pi\rho}
\nonumber \\
& & g  \psi_{\pi\rho} + m_e \psi_e = E \psi_e
\label{coupled_eq}
\ee
These equations are solved to obtain 
the $\pi \rho \to \pi \rho$
scattering $T$ matrix in the form
\be
T_{\pi\rho \to \pi\rho}
&=&
\left( g_R, g \right)
\frac{1}{
\left(
\begin{array}{cc}
g_R(V_{WT}^{-1}- G_{\pi \rho})g_R& g_R G_{\pi \rho} g\\
g G_{\pi \rho} g_R & G_e^{-1} - gG_{\pi \rho} g
\end{array}
\right)
}
\left(
\begin{array}{cc}
g_R\\
g
\end{array}
\right)
\label{Tpirho_sol}
\ee
Here the Green's functions are defined by 
$G_{\pi\rho} = 1/(E - H_{\pi\rho})$ and 
$G_{e} = 1/(E - H_{e})$, 
and $g_R$  the $\pi\rho$ coupling to the 
composite $a_1$ which is defined by the pole 
generated by the Weinberg-Tomozawa interaction $V_{WT}$ as 
\be
T_{WT} = \frac{1}{\displaystyle{{V_{WT}}^{-1}} - G_{\pi \rho}}
\equiv g_R \frac{1}{E-M_{WT}} g_R
\label{T_WT_pole}
\ee
We note that the basis states representing the Hamiltonian  (\ref{H22}) 
are the composite $a_1$ and the elementary one with  
renormalization by the self-energy $gG_{\pi\rho}g$.  
There are some subtleties in, for instance, energy dependence in various coupling 
constants, however.  
Detailed discussions are given in Ref.~\cite{Nagahiro:2011jn}.  

\begin{figure}[h]
\resizebox{0.9\columnwidth}{!}{%
  \includegraphics{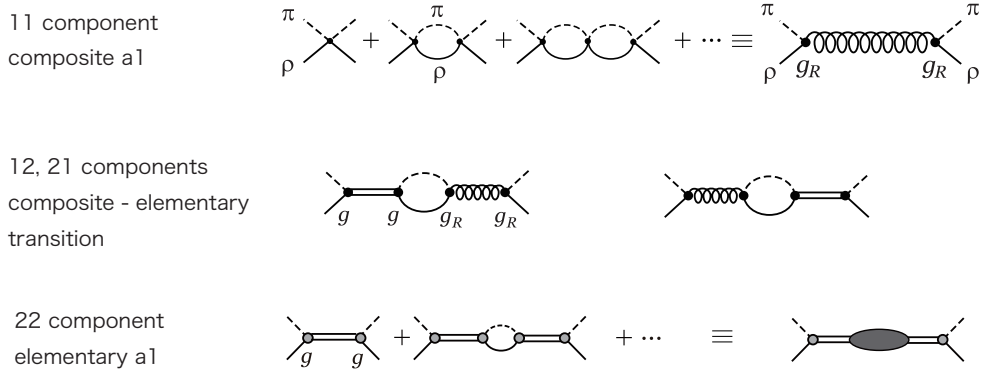} }
\centering
\begin{minipage}{13cm}
\caption{\small Diagrammatic interpretation of various components of the 
effective Hamiltonian, the denominator of (\ref{Tpirho_sol}).}
\label{statediagram}
\end{minipage}
\end{figure}

It is helpful to see the diagrammatic interpretation 
of this scattering matrix  (\ref{Tpirho_sol})
as shown in Fig.~\ref{statediagram}.  
An important point here is that by expressing the $T$ matrix as in (\ref{Tpirho_sol}), 
the physics behind it has become clear; 
the scattering occurs going through resonance states as intermediate states as   
described by the composite and elementary $a_1$'s.  
By diagonalizing the $2 \times 2$ matrix in the denominator 
(\ref{Tpirho_sol}), we obtain the physical poles.  

Various properties of the physical $a_1$'s are discussed in Ref.~\cite{Nagahiro:2011jn}.  
Here we would like to summarize some of them.

\begin{itemize}
\item
When $g \to 0$, the Hamiltonian is just the sum of the decoupled 
components of the $\pi\rho$ system for the composite $a_1$ 
and the elementary $a_1$.  
The locations of these poles are
$M_{1} \sim 1112 - 221 i$ and 
$M_{2} \sim 1189 - 0 i$, 
and determines the model basis for the discussions below.  
The latter is the mass of the $a_1$ given by the original Lagrangian 
(\ref{L_HLS}).  

\item
By varying the coupling strength from zero to the physical value of $g$, 
the two poles deviate from the original locations, and at the physical $g$
they reach
$M_{1} \sim 1033 - 107 i$ and 
$M_{2} \sim 1728 - 313 i$.  
Hence the original composite pole approaches the real axis, while the elementary 
one moves deeply into the imaginary region on the complex energy plane
with much larger real value than the observed physical mass of 
$a_1 \sim 1260$ MeV.  
Therefore, the observed spectrum of $a_1$ is dominantly described by the 
lower pole which is originally of composite nature, though the agreement 
is not perfect.  

\item
At the physical point, the mixing rates of the composite and elementary
components are almost comparative for the pole which is originally composite, 
while  the pole which is originally elementary is also dominated by the composite
component 
(though these 
estimations are only qualitative, because these states appear as resonances and 
the meaning of the mixing ratio is not well defined as in the case of bound states). 
In this way, the physical poles are indeed mixture of the two components, 
and the mixing rates depend on details of the dynamics, in particular 
on hadron dynamics.  

\end{itemize}

We have also studied the resonance properties by changing the color number 
$N_c$~\cite{Nawa:2011pz}.  
We find that at $N_c = 3$, the coupling of the elementary $a_1$ to 
$\pi\rho$ ($g$) and the composite $a_1$ generated by $V_{WT}$ 
play an essential role to 
derive the properties of physical $a_1$, while these terms are of 
higher order in $1/N_c$ and irrelevant in the limit $N_c \to \infty$, 
where only $\bar qq$ mesons survive.  
We have discussed geometrical aspects of the mixing nature in particular 
as function of the color number $N_c$ in Ref.~\cite{Nawa:2011pz}.  

\section{Summary}

In this report, we have discussed hadron resonances starting from a picture of 
quark rearrangement for multiquark systems.  
In general, they develop different configurations  
depending on the properties of the inter-quark forces, a situation
much like atomic nuclei with, for instance, alpha cluster formation.  
As one of most  plausible scenarios, we focus on configurations 
made by two hadronic clusters, a hadronic composite as strong candidate 
of non-standard hadrons which are explained by 
the conventional quark model.  

In addition to well known examples of $\Lambda(1405)$ and sigma mesons, 
we discussed a heavy quark system of exotic quantum number as a strong candidate 
of hadronic composite.  
What is interesting here is that the tensor force plays a crucial role 
just like the mechanism in the deuteron.  
Precisely because of this in the literature a mesonic correspondent, 
for instance $X(3872)$ was called deuson~\cite{Tornqvist:2004qy}. 
We find indeed a bound and resonant states in $J^P = 1/2^-$ and 
$3/2^-$, respectively.  

Finally we have studied mixture of different configurations 
in hadron structure, where an example was shown for 
the axial vector meson $a_1$.  
By solving the two body scattering equation, we have derived a quantum 
mechanical problem of two levels with the basis of the composite and 
elementary $a_1$.  
This analysis implies that there are cases where hadronic components 
become important for the physical hadrons. 

The hadron resonances which are dominated by hadronic composites 
are not yet experimentally confirmed.  
With its loosely bound nature and with various configurations 
than the quark model can provide, further theoretical and experimental 
studies are expected.  
Related to the twin states of $Z_b$ found at Belle~\cite{Collaboration:2011gja}, 
more analogous states are theoretically expected to exist~\cite{Voloshin:2011qa,Ohkoda}.  
Perhaps  interesting features of the strong interaction 
should be seen near the threshold where a new flavor opens and  still 
non-perturbative dynamics  plays important roles to generate hadrons of 
different natures.  

\section*{Acknowledgment}

A.H. is supported in part by the Grant-in-Aid for Scientific
Research on Priority Areas entitled ``Elucidation of New
Hadrons with a Variety of Flavors'' (E01: 21105006).


\end{document}